\documentclass[aps,prb,twocolumn,superscriptaddress,showpacs]{revtex4}

\newcommand{\cH}{\mathcal{H}}

\newcommand{\br}{{\bf r}}
\newcommand{\bk}{{\bf k}}
\newcommand{\bj}{{\bf j}}

\newcommand{\tr}{\,{\rm tr\/}\,}
\newcommand{\lp}{\left(}
\newcommand{\rp}{\right)}
\newcommand{\lb}{\left[}
\newcommand{\rb}{\right]}
\newcommand{\lc}{\left\{}
\newcommand{\rc}{\right\}}

\begin{document}


\title{Hydrodynamic approach to coherent nuclear spin transport}

\author{D. Greenbaum}
\affiliation{ Department of Physics, Massachusetts Institute of
Technology, Cambridge, Massachusetts 02139}
\author{M. Kindermann}
\affiliation{ Department of Physics, Massachusetts Institute of
Technology, Cambridge, Massachusetts 02139}
\author{C. Ramanathan}
\affiliation{Department of Nuclear Engineering, Massachusetts
Institute of Technology, Cambridge, Massachusetts 02139}
\author{D. G. Cory}
\affiliation{Department of Nuclear Engineering, Massachusetts
Institute of Technology, Cambridge, Massachusetts 02139}

\date{\today}

\begin{abstract}
We develop a linear response formalism for nuclear spin diffusion
in a dipolar coupled solid. The theory applies to the
high-temperature, long-wavelength regime studied in the recent
experiments of Boutis {\it et al.} [Phys. Rev. Lett. {\bf 92},
137201 (2004)], which provided direct measurement of interspin
energy diffusion in such a system. A systematic expansion of
Kubo's formula in the flip-flop term of the Hamiltonian is used to
calculate the diffusion coefficients. We show that this approach
is equivalent to the method of Lowe and Gade [Phys. Rev. {\bf
156}, 817 (1967)] and Kaplan [Phys. Rev. B {\bf 2}, 4578 (1970)],
but has several calculational and conceptual advantages. Although
the lowest orders in this expansion agree with the experimental
results for magnetization diffusion, this is not the case for
energy diffusion. Possible reasons for this disparity are
suggested.
\end{abstract}
\pacs{75.45.+j,76.60.-k}
\maketitle

\section{Introduction}

Spin diffusion occurs for conserved quantities in paramagnetic
spin systems.\cite{Bloembergen, Abragam} For dipolar-coupled
nuclear spins in high external magnetic fields, these quantities
are the interaction energy and the spin component along the field.
In a solid, diffusive behavior arises from energy-conserving
two-spin flips of anti-aligned spins, which produce a dynamics
analogous to a random walk.\cite{Bloembergen} Recent experiments
have provided the first direct measurements of
magnetization\cite{Zhang} and energy\cite{Boutis} diffusion in a
dipolar-coupled spin system. These experiments were carried out on
very pure single crystal samples of the insulator calcium
fluoride, a substance that has been studied extensively by the NMR
community for the past sixty years, and whose structural
properties are well known.\cite{Abragam,Goldman} Calcium fluoride
is an ideal material to study because of its very long
spin-lattice relaxation time ($T_1 \sim 100$ s at room temperature
in these experiments), its simple structure (the fluorine
sublattice is simple cubic), and the high accuracy to which the
spin-1/2 nuclei behave as ideal magnetic dipoles. Since the
timescale on which diffusion was observed (on the order of
seconds) was much shorter than $T_1$, the spin system was well
isolated and the evolution coherent. This is the regime in which
Waugh and co-workers demonstrated that spin diffusion is
reversible (by a sign change of an effective Hamiltonian via
multiple-pulse NMR methods).\cite{Waugh} Nevertheless, the
transport of magnetization in three dimensions appears
indistinguishable from an incoherent process, as many calculations
and experiments have shown.\cite{Kh,dG,R,BZ,BW,RY,LG,Kaplan,TW,SW}
In contrast, we find that this does not seem to hold for energy
transport, where despite diffusive behavior the use of thermally
averaged correlation functions leads to results that are
inconsistent with the experiments of Boutis {\em et al}.

This paper is organized as follows. In Secs.\ II and III we
outline the linear response formalism for spin diffusion in a
dipolar-coupled solid, including a derivation of the energy and
magnetization current operators. This formalism applies to the
high-temperature, long-wavelength regime studied experimentally.
It is equivalent to the density matrix approach of Lowe and
Gade,\cite{LG} and Kaplan\cite{Kaplan} (LGK), as we prove in the
Appendix. Our formulation has the advantages that the
long-wavelength limit is built-in and the application to
inter-spin energy is straightforward. In Sec.\ IV, we derive the
expansion of the Kubo formula in powers of the flip-flop term of
the dipolar Hamiltonian, which we use for numerical evaluation of
the diffusion coefficients. The diffusion coefficients of
magnetization and inter-spin energy obtained from this expansion
to two leading orders in the flip-flop are given in Sec.\ V. Their
numerical values are calculated in Sec.\ VI, along with an
estimate of the errors. Our main finding is that, although the
expansion for the magnetization diffusion coefficient reproduces
the experimental results, the series for inter-spin energy does
not. In Sec.\ VII we discuss this result and comment on how our
method complements previous approaches to the problem.

\section{Kubo Formula}

The linear response diffusion coefficients of a continuous system
may be obtained from Kubo's formula,\cite{Forster}
\begin{equation}
D = \frac{\int_0^\infty dt \int d^3 \br \int d^3 \br' \langle
j_z(\br,t) j_z(\br',0) \rangle_{eq}}{\int d^3 \br \int d^3 \br'
\langle S(\br,0) S(\br',0) \rangle_{eq}}, \label{Kubo1}
\end{equation}
where $S(\br)$ is the operator representing the local density of a
globally conserved quantity. In our case this is either the energy
($S = \cH$) or the component of spin magnetization along an
external field ($S=M$), and $j_z(\br)$ is the corresponding
current density along the transport direction. The angular
brackets denote averaging over an equilibrium statistical ensemble
at the appropriate temperature, which is infinite for the case we
consider. The discrete version of this formula is more directly
applicable to our problem since the spins reside on a lattice, and
is given by replacing $\br$ with a lattice site index. However,
the continuum representation is more useful for derivations, and
is completely equivalent.

Eq.\ (\ref{Kubo1}) is the correct formula for the diffusion
coefficient of the quantity $S$ if the following assumptions hold.
\begin{description}
\item[(1)] The correlation function of $S$ is known to have a
diffusive form (i.e. a diffusive pole). \item[(2)] The system may
at all times be described by a statistical ensemble that is
sufficiently close to equilibrium (linear response regime).
\end{description}
Assumption (1) may be checked experimentally, and has been
verified for both spin-spin and energy-energy correlators of
dipolar-coupled spins in a solid.\cite{Zhang,Boutis} Assumption
(2) is more subtle, as it rests on the validity of the ergodic
hypothesis, which has received much attention recently in the
context of lattice spin systems in dimensions 1 --
3.\cite{Waugh2,FBE,FL,PU,Fine} As discussed below, we believe that
it may not be valid for the diffusion of energy.

Experimentally, spin diffusion has been measured by observing the
relaxation of initial states varying sinusoidally in real space
with a given wavevector $\bk$.\cite{Zhang,Boutis} Such
spatially-inhomogeneous states are represented mathematically as
perturbations on the infinite temperature equilibrium state,
$\rho_\infty = {\bf 1}/2^N$, where $N$ is the number of spins. We
have
\begin{equation}
\rho(0) = \rho_\infty + \delta\rho(0). \label{rho}
\end{equation}

Two possibilities exist for $\delta\rho(0)$, corresponding to
long-wavelength fluctuations in the two conserved quantities,
\begin{eqnarray}
\delta\rho_M(\bk,0) = \epsilon \int d^3 \br \cos(\bk\cdot\br)
M(\br), \label{drz} \\
\delta\rho_\cH(\bk,0) = \epsilon \int d^3 \br \cos (\bk\cdot\br)
\cH(\br), \label{drd}
\end{eqnarray}
where $\epsilon$ is a small quantity of order $\gamma\hbar B_0 /
2^N k_B T$. Here $B_0$ is the external field, and $\gamma$ is the
gyromagnetic ratio of the nuclear species of interest ($\gamma =
2.51\times 10^{4} \; {\rm rad \; Hz / Oe}$ for $^{19}$F). Using
typical experimental values of $B_0 = 1\ {\rm T},\; T = 300\ {\rm
K}$, we estimate $\epsilon \sim 10^{-5} - 10^{-6}$. The important
length scales are related by $L
>> k^{-1}
>> a$, where $L$ is the sample size and $a$ is the lattice
spacing. $a = 2.73\times 10^{-8}\,{\rm cm}$ for the Fluorines in
CaF$_2$, and $L \sim 0.1\,{\rm cm}$, $k^{-1} \sim 10^{-4}\,{\rm
cm}$ in the experiments. The realization of such states is
discussed in detail by Boutis {\em et al}.\cite{Boutis} In
practice, $\bk$ is parallel to the external magnetic field, and
transport is measured in the same direction. As shown in the
Appendix, expectation values taken with respect to the states in
Eqs.\ (\ref{drz}) and (\ref{drd}) give rise to the
equilibrium-averaged correlation functions appearing in Eq.\
(\ref{Kubo1}).

We find it plausible that assumption (2) is valid when the spatial
profile of magnetization or energy in the initial state varies
sufficiently slowly, and when this state has no long range
correlations. While this is possible for magnetization, the
creation of spatial inhomogeneities in the interaction energy
requires NMR techniques that introduce short - ranged correlations
between the spins. Some of these correlations persist due to
energy conservation, and these are the ones relevant to energy
transport. Other correlations are assumed to decay rapidly and to
be unobservable. A careful analysis of the initial state involved
in the energy diffusion measurements is needed to determine
whether assumption (2) is satisfied. On the other hand, the extent
to which our results agree with experiment may be viewed as a
constraint on its validity.

\section{Current Operators}

In order to use Eq.\ (\ref{Kubo1}) we need to obtain expressions
for the current operators $j_z(\br)$. We start with the rotating
-- frame Hamiltonian of interacting magnetic dipoles in a large
external magnetic field.\cite{Slichter, Abragam}
\begin{eqnarray}
\cH =&& \sum_{i,j(i\neq j)} \cH_{ij}, \label{h1} \\
\cH_{ij} =&& b_{ij} \lp I_{iz} I_{jz} - \frac{1}{4}\left(I_{i+}
I_{j-} + I_{j+} I_{i-} \right) \rp. \label{h2}
\end{eqnarray}
The latin indices run over all lattice sites and the $I_{i\alpha}$
are operators for spin-1/2, satisfying the commutation relations
$[I_{i\alpha},I_{j\beta}] = \delta_{ij}I_{i\gamma}$, where
$\alpha,\,\beta,\,\gamma$ is any cyclic permutation of x, y, z.
The $I_{j\pm} \equiv I_{jx} \pm iI_{jy}$ are raising and lowering
operators. The coupling constants are given by
\begin{equation}
b_{ij} = \frac{\gamma^2 \hbar^2}{2} \frac{1 - 3
\hat{\br}_{ij}\cdot\hat{\bf z}}{r_{ij}^3}, \label{bij}
\end{equation}
where $\br_{ij}$ is the displacement vector between lattice sites
$i$ and $j$, and unit vectors are denoted by a caret. We take the
z-direction to be along the external magnetic field.

The total $z$-component of spin and total energy are conserved
quantities of the Hamiltonian, Eqs.\ (\ref{h1}) and (\ref{h2}).
This implies that they satisfy a local continuity
equation.\cite{Forster} We define spin and interaction energy
densities that depend on continuous variables,\cite{BZ,FG}
\begin{eqnarray}
I_{\alpha}({\bf r},t) \equiv&& \sum_{i} \delta({\bf r - r}_i)
I_{i\alpha}(t), \;\;\; \alpha = z, +, - \label{spin-density} \\
\cH({\bf r},t) \equiv&& \sum_{i,j (i \neq j)}\delta({\bf r -
r}_i)\cH_{ij}(t) \label{energy-density}.
\end{eqnarray}
The operators in these equations and throughout this section are
in the Heisenberg representation.

The densities of magnetization and interaction energy are given in
terms of the spin densities as
\begin{eqnarray}
M(\br) = &&\gamma\hbar I_{iz}(\br), \\
\cH(\br) = &&\int d^3\br' b(\br - \br') \{ I_z(\br) I_z(\br')
\nonumber \\ &&- \frac{1}{4}\lb I_+(\br) I_-(\br') + I_+(\br')
I_-(\br) \rb\}, \label{H(r)}
\end{eqnarray}
%
where $b(\br-\br') = \sum_{i,j}\delta({\bf r - r}_i)\delta({\bf r'
- r}_j)b_{ij}$. We avoid writing the time dependence of the
operators where there is no possibility of confusion. Next, we use
the Heisenberg equation of motion,
\begin{eqnarray}
\frac{\partial M({\bf r},t)}{\partial t} =&&
-\frac{i}{\hbar}[M({\bf r},t), \cH], \label{Heisenberg-spin} \\
\frac{\partial \cH(\br,t)}{\partial t} =&&
-\frac{i}{\hbar}[\cH(\br,t),\cH], \label{Heisenberg-energy}
\end{eqnarray}
to obtain
\begin{eqnarray}
\frac{\partial M({\bf r})}{\partial t} =&& \frac{i\gamma\hbar}{2}
\int d^3 {\bf r'}b({\bf r - r'})\lp I_{+}({\bf r}) I_{-}({\bf r'})
\right.\nonumber \\ && \left. \hspace*{1.3in}
- I_{+}({\bf r'}) I_{-}({\bf r})\rp, \nonumber\\\label{dIdt} \\
\frac{\partial \cH(\br)}{\partial t} =&& \frac{i}{4}\int
d^3{\br'}d^3{\br''}b(\br' - \br)b(\br''-\br') \nonumber \\&&
\hspace*{-0.2in}\times \{I_z(\br')\lp I_+(\br)I_-(\br'') -
I_+(\br'')I_-(\br)\rp \nonumber \\&& \hspace*{-0.2in}+
2I_z(\br)\lp I_+(\br')I_-(\br'') - I_+(\br'')I_-(\br')\rp
\nonumber \\&& \hspace*{-0.2in}+ 2I_z(\br'')\lp I_+(\br)I_-(\br')
- I_+(\br')I_-(\br)\rp\}. \label{dHdt}
\end{eqnarray}
In order to write Eqs.\ (\ref{dIdt}) and (\ref{dHdt}) as
continuity equations in the usual form, they may be integrated
over an arbitrary volume and the result expressed in terms of a
surface integral, whose integrand is the current density. This
gives the current density operators,
\begin{eqnarray}
\bj^{(M)}(\br,t) =&& \frac{i\gamma\hbar}{4}\int d^3 {\bf r'}b({\bf
r - r'})(\br - \br')\nonumber \\ && \times\lb I_{+}({\bf r})
I_{-}({\bf r'}) - I_{+}({\bf r'})
I_{-}({\bf r})\rb, \label{mag-current}\\
\bj^{(\cH)}(\br,t) =&& \frac{i}{8}\int d^3{\br'}d^3{\br''}b(\br' -
\br)b(\br''-\br')(\br - \br'') \nonumber \\ &&
\hspace*{-0.2in}\times
\{I_z(\br')\lb I_+(\br)I_-(\br'') - I_+(\br'')I_-(\br)\rb \nonumber \\
&& \hspace*{-0.2in}+ 2I_z(\br)\lb I_+(\br')I_-(\br'') -
I_+(\br'')I_-(\br')\rb \nonumber \\ &&\hspace*{-0.2in} +
2I_z(\br'')\lb I_+(\br)I_-(\br') - I_+(\br')I_-(\br)\rb\},
\label{energy-current}
\end{eqnarray}
where the superscripts $M$ and $\cH$ denote magnetization and
energy current, respectively. The continuity equations then take
their usual form,
\begin{equation}
\frac{\partial S(\br,t)}{\partial t} + {\bf \nabla} \cdot
\bj^{(S)}(\br,t) = 0, \;\;\; S = M,\cH. \label{continuity}
\end{equation}
Similar results have been obtained by Furman and Goren by a
different method.\cite{FG,Furman}

\section{Perturbation Theory}

Using an interaction representation introduced by Lowe and
Norberg,\cite{LN} we can expand Eq.\ (\ref{Kubo1}) in powers of
the flip-flop term of the Hamiltonian. Following Ref.\ 13 we
define
\begin{eqnarray}
&&\cH_1 = \sum_{i,j(i\neq j)}b_{ij} \, I_{iz}I_{jz}, \label{H1}\\
&&\cH_2 = -\frac{1}{4}\sum_{i,j(i\neq j)}b_{ij} \, (I_{i+}I_{j-} +
I_{i-}I_{j+}). \label{H2}
\end{eqnarray}
Using the notation $\tilde{A} (t) = e^{-i\cH_1t/\hbar} A(0)
e^{i\cH_1t/\hbar}$, we may write any operator in the Heisenberg
representation as an infinite series in $\cH_2$,
\begin{eqnarray}
&&A(t) = e^{i\cH_1t/\hbar}\lc A(0) + \frac{i}{\hbar}\int_0^t dt_1
\; [\tilde{\cH}_2(t_1),A(0)] \right.
\nonumber \\
&& + \lp \frac{i}{\hbar} \rp ^2\int_0^t \ dt_1 \ \int_0^{t_1} \
dt_2\,
[\tilde{\cH}_2(t_1),[\tilde{\cH}_2(t_2),A(0)]] \nonumber \\
&& + \left.\vphantom{\int_0^t} \cdots \rc e^{-i\cH_1t/\hbar}.
\end{eqnarray}
Using this expansion for $j_z(t)$ in Eq.\ (\ref{Kubo1}) gives a
perturbation series for the diffusion coefficient.
\begin{eqnarray}
&&D = \lp\int_0^\infty \, dt\langle j_z(0)\tilde{j}_z(t)\rangle \right.\nonumber \\
&&+ \frac{i}{\hbar}\int_0^\infty\,dt\,\int_0^t\, dt_1\langle
[\tilde{\cH}_2(t_1),j_z(0)]\tilde{j}_z(t)\rangle \nonumber \\
&&+ \lp \frac{i}{\hbar} \rp ^2\int_0^\infty dt\int_0^t  dt_1
\int_0^{t_1}  dt_2 \langle
[\tilde{\cH}_2(t_1),[\tilde{\cH}_2(t_2),j_z(0)]] \nonumber
\\
&&\left. \times \tilde{j}_z(t)\rangle + \cdots
\vphantom{\int_0^t}\rp/\langle S(0)^2\rangle, \label{D}
\end{eqnarray}
where the spatial integrations have been suppressed, and angular
brackets denote an equilibrium average. The operators
$\tilde{\cH}_2(t)$ and $\tilde{j}_z(t)$ can be evaluated using the
identity,\cite{LN,LG}
\begin{eqnarray}
&\exp&\hspace{-2ex}\lp it\hspace{-2ex}\sum_{m,n (m \neq
n)}\hspace{-2ex} b_{mn} I_{mz} I_{nz}\hspace{-1ex}\rp
\hspace{-1ex}I_{i+} I_{j-}\hspace{-1ex}\exp \lp
-it\hspace{-2ex}\sum_{m,n (m \neq
n)}\hspace{-2ex}b_{mn} I_{mz} I_{nz}\hspace{-1ex}\rp \nonumber \\
&&= I_{i+} I_{j-} L_{ij}(t), \label{LN-identity}
\end{eqnarray}
%
where
\begin{equation}
L_{ij}(t) \equiv \prod_{l (l \neq i,j)} e^{2it(b_{li} - b_{lj})
I_{lz}}. \label{Lij}
\end{equation}
%
It is advantageous to approximate the operator $L_{ij}(t)$ by a
c-number, equal to its thermal average, $L_{ij}(t) \approx
G_{ij}(t)$. At infinite temperature,
\begin{equation}
G_{ij}(t) = \prod_{m (m\neq i,j)} \cos(b_{im} - b_{jm})t.
\label{G}
\end{equation}
The neglected q-number terms are expected to be approximately
$20-25\%$ smaller, as discussed at the end of Sec.\ IV.

It turns out that $G_{ij}(t)$ depends very weakly on its indices.
In order to make the calculation of higher order terms in the
perturbation series tractable, we therefore replace $G_{ij}(t)$ by
$G(t)$. $G$ is a suitably defined average over all the $G_{ij}$'s,
as discussed in Sec.\ VI. The overall error introduced by
approximating the higher order terms in this way is therefore
small (we shall see in Sec.\ VI that it is less than 10\%). The
resulting perturbation series contains only even-order terms,
\begin{eqnarray}
&&D = \sum_{n=0}^\infty \lp\frac{1}{\hbar}\rp^{2n}
\frac{F^{2n+1}}{(2n+1)!}\frac{\langle [\cH_2,j_z(0)]_n^2\rangle}{\langle S(0)^2 \rangle},\label{D-series}\\
&&F \equiv \int_0^{\infty} G(t)\,dt, \label{F}
\end{eqnarray}
where $[A,B]_n \equiv [A,[A,...,[A,B]...]]$ is a commutator with
$A$ taken $n$ times. It is worth noting that this series may be
written in a simple closed form,
\begin{eqnarray}
D = \frac{ \int_0^{F}dF'\langle e^{i\cH_2F'/\hbar} j_z(0)
e^{-i\cH_2F'/\hbar} j_z(0)\rangle}{\langle S(0)^2\rangle},
\end{eqnarray}
which shows that our system is approximately equivalent to one
with purely flip-flop interaction, evolving for a finite time, F.

\section{Analytical Results}

In this section we give analytic expressions for the diffusion
coefficients to the two leading orders using Eq.\ (\ref{D}) and
(\ref{D-series}). We are interested in the infinite temperature
case. Therefore, $\langle \cdots \rangle_{eq} =
\tr\{\cdots\}/2^N$, where $N$ is the number of lattice sites. Each
term in Eq.\ (\ref{D-series}) then contains a trace over a product
of spin operators at different lattice sites. The evaluation of
such traces is a tedious but straightforward task. We do not
discuss it further here.

\subsection{Magnetization Diffusion}

The denominator of Eq.\ (\ref{Kubo1}) for $S = M$ at $T = \infty$
is $\int d^3 \br \int d^3 \br' \tr\{ M(\br,0) M(\br',0) \}/2^N =
N\gamma^2\hbar^2 /4$.

The lowest order term is calculated exactly, by inserting the
expressions for the Hamiltonian, Eq.\ (\ref{H1}), and the
magnetization current, Eq.(\ref{mag-current}), into the first line
of Eq.\ (\ref{D}), and applying the identity, Eq.\
(\ref{LN-identity}). We obtain
\begin{eqnarray}
D_M^{(0)} =&& \frac{1}{4}\sum_{i (i\neq 0)} b_{0i}^2 z_{0i}^2
F_{0i}, \label{DM0} \\
z_{ij} \equiv && z_i - z_j,
\end{eqnarray}
where $F_{ij}$ is given by Eq.\ (\ref{F}) with $G_{ij}(t)$
replacing $G(t)$. The lattice site labelled $0$ is arbitrary, due
to the translational invariance of an infinite lattice. This
result is identical to LGK's, as expected from the equivalence of
the two methods (see Appendix).

The next term, obtained from Eq.\ (\ref{D-series}) using all of
the approximations discussed in the last section, is
\begin{equation}
D_M^{(2)} = \frac{1}{8}\frac{F^3}{3!}\lp 2\sum_{i (i\neq 0)}
b_{0i}^4 z_{0i}^2 - \sum_{i,j(i,j\neq0)} b_{0i}^2b_{0j}^2z_{0j}^2
\rp. \label{DM2}
\end{equation}

\subsection{Energy Diffusion}

For $S = \cH$, the denominator is $\int d^3 \br \int d^3 \br'
\tr\{ \cH(\br,0) \cH(\br',0) \} = (3/16)\times 2^N\times\sum_{i,j
(i\neq j)} b_{ij}^2$. The lowest order term is
%
\begin{eqnarray}
D_\cH^{(0)} &=&\frac{1}{48\left(\sum_{i,j(i\neq
j)}b_{ij}^2\right)}\lp\sum_{i,j,k (i\neq j\neq k)}
B_{ijk}^2F_{ik}\right. \nonumber \\ &&\left. -
2\hspace{-5ex}\sum_{i,j,k,l (i\neq j\neq k\neq l)} B_{ijk}B_{ilk}
R_{ik;jl} \rp,
 \label{DH0}
\end{eqnarray}
where
\begin{eqnarray}
B_{ijk} \equiv && b_{ij}b_{jk}z_{ik} + 2b_{ij}b_{ik}z_{jk} +
2b_{ik}b_{jk}z_{ij}. \nonumber \\ R_{kl;ij} \equiv &&
\int_0^\infty dt K_{kl;ij}(t), \label{R}
\\ K_{kl;ij}(t) &\equiv& -4\tr\{I_{iz}I_{jz}L_{kl}(t)\}|_{i \neq j \neq k \neq l}
\nonumber \\ &=&\hspace{-1ex}
\sin(\Delta_{kl;i}t)\sin(\Delta_{kl;j}t)\hspace{-3ex}\prod_{p
(p\neq
i,j,k,l)}\hspace{-3ex}\cos(\Delta_{kl;p}t), \label{K} \\
\Delta_{kl;p} \equiv && b_{pk} - b_{pl}.
\end{eqnarray}
%

It is useful to have approximate analytic expressions for the
integrals in Eq.\ (\ref{F}) and Eq.\ (\ref{R}). It has been shown
that the saddle point approximation to $F_{ij}$ is quite
accurate.\cite{LG,Kaplan} It is equivalent to replacing $G_{ij}$
in the integrand by a Gaussian,
\begin{equation}
G_{ij}(t) \approx \exp\lp-\frac{1}{2}\sum_{k (k\neq
i,j)}(b_{ki}-b_{kj})^2t^2\rp. \label{Gsp}
\end{equation}
To evaluate $R_{kl;ij}$, we replace the product of cosines in
$K_{kl;ij}(t)$ by a Gaussian, as above. Since this Gaussian cuts
off the integral at times $t \ll \hbar/b$, where $b$ is of the
order of the nearest neighbor coupling strength, we can expand the
sine terms around $t=0$, which yields
\begin{eqnarray}
K_{kl;ij}(t) &&\approx (b_{ik}-b_{il})(b_{jk}-b_{jl})t^2 \nonumber
\\ &&\times\exp\lp-\frac{1}{2}\sum_{p (p\neq
i,j,k,l)}(b_{pk}-b_{pl})^2t^2\rp, \label{Ksp}
\end{eqnarray}
These approximations give
\begin{eqnarray}
F_{ij} = \frac{\sqrt{\pi}}{\sqrt{2\sum_{k (k\neq
i,j)}(b_{ki}-b_{kj})^2}}, \label{Fsp} \\
R_{kl;ij} =
\sqrt{\frac{\pi}{2}}\frac{(b_{ik}-b_{il})(b_{jk}-b_{jl})}{\lp
\sum_{p (p\neq i,j,k,l)}(b_{pk}-b_{pl})^2\rp^{3/2}}. \label{Rsp}
\end{eqnarray}
Equation (\ref{Fsp}) has been derived previously.\cite{LG,Kaplan}
Equations\ (\ref{Fsp}) and (\ref{Rsp}) allow a more rapid
numerical evaluation of the expressions for the diffusion
coefficients than by numerical integration of Eqs.\ (\ref{G}) and
(\ref{K}). By numerical integration on cubic lattices of between
$5^3$ and $11^3$ spins, we have verified that Eq.\ (\ref{Fsp})
approximates the exact value with an error that is less than 2\%
and Eq.\ (\ref{Rsp}) approximates the exact value to within 10\%.

We found numerically that the sum over $R$ terms in Eq.\
(\ref{DH0}) was about $20-25\%$ in magnitude of the sum over $F$
terms, for both the (001) and (111) directions. The $R$ terms
arise from the lowest order (in $I_{iz}$ operators) q-number
correction to the approximation, Eq.\ (\ref{G}), to $L_{ij}(t)$.
We therefore expect this same number to also be a good estimate of
the error in this approximation.

Using Eq.\ (\ref{D-series}) for the next order term, we obtain
\begin{eqnarray}
 D_\cH^{(2)} &=& \frac{F^3}{(3!) (192) \sum_{k,l (k\neq
l)}b_{kl}^2} \nonumber \\ &&\times\lp \sum_{u,q,l (u\neq q \neq
l)}\hspace{-4ex}\lp 4b_{uq}^2 B_{ulq}^2\hspace{-0.5ex}+\hspace{-0.5ex}8b_{uq}^2 B_{qul}^2\hspace{-0.5ex}-\hspace{-0.5ex}4b_{uq}b_{ql}B_{qul}B_{ulq}\rp \right. \nonumber \\
&&+ \hspace{-5ex}\sum_{u,q,l,k (u\neq q \neq l \neq k)} \lb
b_{uq}^2 \lp -3B_{kql}^2 - B_{qkl}^2 + 2B_{kql}B_{kul}\rp \right.
\nonumber \\&&+ b_{uq}b_{kq}\lp -4B_{uql}B_{kql} + 6B_{ulq}B_{qlk}
+ 3B_{ukl}B_{kql}\right.\nonumber\\&&\left.-4B_{kul}B_{ukl} \rp
+b_{uq}b_{kl}\lp 4B_{kul}B_{qku} + 2B_{quk}B_{ukl}\right.
\nonumber\\&&\left.\left.\left.-2B_{kqu}B_{qkl} -
4B_{qku}B_{kql}\rp \rb \vphantom{\sum_{u,q,l (u\neq q \neq
l)}}\rp, \label{DH2}
\end{eqnarray}

\section{Numerical Results}
Because Eqs.\ (\ref{DM0}), (\ref{DM2}), (\ref{DH0}), and
(\ref{DH2}) can be evaluated numerically only for finite lattice
sizes, we use finite size scaling to extract the infinite lattice
limit. The approach to the infinite lattice value is expected to
follow a power law. For example, if we approximate the sums by
integrals in Eq.\ (\ref{DM0}),
\begin{eqnarray}
D_M^{(0)} \approx \frac{1}{4}\int_{a \leq r \leq L}d^3\br \,
b(\br)^2z^2F(\br).
\end{eqnarray}
As $r \rightarrow\infty$, $F \rightarrow {\rm const}$. Therefore,
\begin{eqnarray}
D_M^{(0)} \sim &&{\rm const} \times \int_a^L r^2dr \lp
\frac{1}{r^3}\rp^2r^2 \nonumber \\
= &&{\rm const} \times \lp \frac{1}{a} - \frac{1}{L} \rp.
\end{eqnarray}
A least squares fit to a power law describes very well the scaling
with finite lattice size. For diffusion of magnetization, we were
able to vary the lattice size, in increments of 2 lattice sites,
between 1 and 81 lattice sites on an edge. For energy diffusion,
we studied lattices with up to 27 sites on an edge.

The constant $F$ in Eqs.\ (\ref{DM2}) and (\ref{DH2}) was taken as
the mean value of the $F_{ij}$ over three layers of nearest
neighbors. This averaging procedure works well since contributions
from far-apart indices are suppressed by the $b$ factors in the
summations. In dimensionless units, this gave
\begin{equation}
\frac{\gamma^2\hbar}{a^3} \times F = \lc \begin{array}{cc} 0.48 \pm 0.05, & (001)\;{\rm direction},  \\
1.17 \pm 0.14, & (111)\;{\rm direction}. \end{array} \right.
\label{Fconst}
\end{equation}
The error, taken to be one standard deviation from the mean, is
$10\%$ for the (001) direction and $12\%$ for the (111) direction.
Therefore, the error in $F^3$ is about three times as much, or
between $30\%$ and $40\%$. As the magnitude of the second order
terms is between $10\%$ and $20\%$ that of the zero order terms,
the overall error in approximating $F_{ij} \approx F$ is less than
$10\%$.

The results, including errors due to fitting and approximations,
are summarized in Table \ref{tableofresults}. Our magnetization
diffusion coefficient agrees well with experimental values.
However, the energy diffusion coefficient that we obtain is
significantly smaller than observed experimentally, albeit larger
than the magnetization diffusion coefficient. The relative
orientation dependence is within the experimental range for
magnetization, but disagrees drastically for energy.

\begin{table}
\caption{Summary of the theoretical and experimental results for
the spin diffusion rate of energy, $D_\cH$, and magnetization,
$D_M$ for a single crystal of calcium fluoride. Theoretical values
have been obtained by numerically evaluating Eqs.\ (\ref{DH0}) and
(\ref{DH2}), and Eqs.\ (\ref{DM0}) and (\ref{DM2}), using Eq.\
(\ref{Fsp}) for $F_{ij}$ and Eq.\ (\ref{Rsp}) for $R_{ikmj}$, and
using finite size scaling to extrapolate to the infinite lattice
limit.} \hspace*{-0.05in}\begin{tabular} {||c|c|c|c||} \hline
Theory&[001]&[111]&$D_{001}/D_{111}$  \\ \hline
\hspace*{0.1in} $D_{M}^{(0)}$ \hspace*{0.1in} ($ \times 10^{-12}$cm$^{2}$/s) & $8.4 \pm 0.2$ & $7.9 \pm 0.2$ & -- \\
\hspace*{0.1in} $D_{M}^{(2)}$ \hspace*{0.1in} ($ \times 10^{-12}$cm$^{2}$/s) & $-0.3 \pm 0.1$ & $-0.5 \pm 0.2$ & -- \\
\hline $D_M^{(0+2)}$\hspace*{0.1in} ($ \times 10^{-12}$cm$^{2}$/s)
& $8.1 \pm 0.3$ & $7.4 \pm 0.4$ & $1.1 \pm 0.1$ \\ \hline
\hspace*{0.1in} $D_{\cH}^{(0)}$ \hspace*{0.1in} ($ \times 10^{-12}$cm$^{2}$/s) & $19.8 \pm 0.7$ & $12.5 \pm 0.3$ & --  \\
\hspace*{0.1in} $D_{\cH}^{(2)}$ \hspace*{0.1in} ($ \times 10^{-12}$cm$^{2}$/s) & $-0.7 \pm 0.3$ & $-1.1 \pm 0.5$ & --  \\
\hline $D_\cH^{(0+2)}$\hspace*{0.1in} ($ \times
10^{-12}$cm$^{2}$/s) & $19.1 \pm 1.0$ & $11.4 \pm 0.8$ & $1.7 \pm 0.2$\\
 \hline Experiment, $D_M$ & [001] & [111]& $D_{001}/D_{111}$  \\ \hline
\hspace*{0.1in} Ref. \cite{Zhang} $(\times 10^{-12}$ cm$^2$/s) & $7.1 \pm 0.5$ & $5.3 \pm 0.3$ & $1.34 \pm 0.12$\\
\hspace*{0.1in} Ref. \cite{Boutis} $(\times 10^{-12}$ cm$^2$/s) &
$6.4 \pm 0.9$ & $4.4 \pm 0.5$ & $1.45 \pm 0.26$\\ \hline
Experiment, $D_\cH$ & [001] & [111]& $D_{001}/D_{111}$
\\ \hline \hspace*{0.1in} Ref. \cite{Boutis} $(\times 10^{-12}$
cm$^2$/s) & $29 \pm 3$ & $33 \pm 4$ & $0.88 \pm 0.14$\\ \hline
\end{tabular}
\label{tableofresults}
\end{table}

\section{Conclusion}
We have presented a hydrodynamic approach to study the
long-wavelength spin dynamics in a lattice of dipolar-coupled
spins-1/2 in high magnetic field and at high temperature,
motivated by recent experiments on coherent nuclear spin transport
in calcium fluoride. The Kubo formula, Eq.\ (\ref{Kubo1}), for the
diffusion coefficients applies to the physical regime probed
experimentally and rests on the assumption that the time evolution
of the system is ergodic. We developed a perturbation theory for
Eq.\ (\ref{Kubo1}) that is equivalent to the approach of LGK but
simplifies the calculations enormously. This allowed us to obtain
the diffusion coefficients for magnetization as well as energy to
leading order in the flip-flop term of the Hamiltonian, and
estimate the first perturbative correction. The result for
magnetization diffusion agrees with experiment to within its
degree of accuracy. The result for inter-spin energy diffusion is
larger than that for magnetization diffusion, in qualitative
agreement with the experiment. It does not, however, describe the
experiment quantitatively.

One possible reason for the disparity is the ergodicity
assumption, which is implicit in our choice of how to take the
average in the correlation functions appearing in Eq.\
(\ref{Kubo1}). Our assumption of Eq.\ (\ref{rho}) leads to the
equivalence of the non-equilibrium average of the conserved
density to its equilibrium correlation function, as is usual in
linear response theory.\cite{Forster} For magnetization diffusion,
the low polarization results in a very low density of polarized
spins in a completely randomly polarized background. The sparsity
of polarized spins means that their effect on each other is
negligible, and they can be treated as independent. This suggests
that statistical averaging over a complete infinite temperature
ensemble is the correct procedure. For energy diffusion, the
correlations inherent to the initial states used in the
experiments of Boutis {\em et al.} may require the use of an
ensemble that is a subset of the full Hilbert space. This would
imply a modification of Eq.(\ref{rho}) for the density matrix at
finite time, and needs to be investigated further.

The most evident source of inaccuracy in our calculation is our
truncation of the perturbation series. Our estimate of the
next-to-leading order correction to energy diffusion does not rule
out the possibility that a resummation of our perturbation series
would explain the experiment quantitatively. However, currently
available non-perturbative (e.g. Bennet and Martin\cite{BM}) and
resummation (e.g. Borckmans and Walgraef\cite{BW}) methods are
much too cumbersome to treat the diffusion of inter-spin energy.
They are also plagued by their own uncontrolled approximations,
such as the ad-hoc replacement of certain correlation functions by
gaussians in order to simplify the calculations. This state of
affairs, coupled with the great success of LGK at predicting the
magnetization diffusion coefficient based on an expansion in the
flip-flop term, and earlier that of Lowe and Norberg\cite{LN} at
fitting the shape of the FID by a similar expansion, motivated us
to systematize their approach and apply it to inter-spin energy
diffusion. For the final word on this problem, we await either a
new non-perturbative method or a tractable technique for summing
our perturbation series to all orders.

\begin{acknowledgements} We would like to thank N. Boulant, P.
Cappellaro, H. Cho, J. Hodges, D. Pushin and Professor L. S.
Levitov for discussions, and the NSF, DARPA, ARO, ARDA, and CMI
for financial support.
\end{acknowledgements}

\appendix*

\section{Equivalence to LGK Method}
LGK compute the time evolution of the average density, $\langle
S(\bk,t) \rangle = \tr\{\delta\rho(-\bk,t)S(\bk)\}$, starting from
one of the non-equilibrium states, Eq.\ (\ref{drz}) or Eq.\
(\ref{drd}). In general, we write
\begin{equation}
\delta\rho(t=0) = \epsilon \int d^3 \br \cos(\bk\cdot\br)S(\br) =
\frac{\epsilon}{2}S(\bk) + \frac{\epsilon}{2}S(-\bk).
\end{equation}
The average of any Schr\"odinger operator $A$ at time $t$ is
\begin{equation}
\langle A(t) \rangle = \tr\{\delta\rho(t)A\}.
\end{equation}
If $\langle S(\br,t) \rangle$ satisfies a diffusion equation,
\begin{equation}
\frac{\partial}{\partial t}\langle S(\br,t) \rangle = D \nabla^2
\langle S(\br,t) \rangle,
\end{equation}
then the time dependence of $\langle S(\bk,t) \rangle$ [the
Fourier transform of $\langle S(\br,t) \rangle$ at wavevector
$\bk$] is given by
\begin{eqnarray}
\frac{\partial}{\partial t} \langle S(\bk,t) \rangle =&&
-Dk^2\langle S(\bk,t) \rangle \nonumber \\ \Longrightarrow
\hspace{0.2in} \langle S(\bk,t) \rangle =&& e^{-k^2Dt}\langle
S(\bk,0) \rangle.
\end{eqnarray}
Taking the time derivative of the last equation and rearranging
terms, we obtain LGK's expression for the diffusion coefficient,
\begin{equation}
D = \lim_{k \rightarrow 0}
\left(-\frac{1}{k^2}\right)\frac{\langle \dot{S}(\bk,t)
\rangle}{\langle S(\bk,0) \rangle}.\label{LGK}
\end{equation}
The right hand side is in fact independent of time for $t$ greater
than the short timescale defined by the inverse spin-spin
coupling. We keep the explicit time dependence to remind us that
$t$ is long, but finite, while $k$ tends to zero. In other words,
we take the $t \rightarrow \infty$ limit after the $k \rightarrow
0$ limit.

In the Heisenberg representation, we have
\begin{eqnarray}
\langle S(\bk,t) \rangle =&& \tr\{e^{-i\cH t}\delta\rho(0)e^{i\cH
t} S(\bk)\} \nonumber \\=&&
\frac{\epsilon}{2}\tr\{S(\bk,-t)S(\bk,0)\} \nonumber \\
&& + \frac{\epsilon}{2}\tr\{S(-\bk,-t)S(\bk,0)\}. \label{eq}
\end{eqnarray}
Because of translational invariance, only the second term on the
right side of Eq.\ (\ref{eq}) contributes, and we have
\begin{equation}
\langle S(\bk,t) \rangle =
\frac{\epsilon}{2}\tr\{S(-\bk,-t)S(\bk,0)\}. \label{Sk}
\end{equation}
We take the time derivative of Eq.\ (\ref{Sk}), and use $\partial
S(t)/\partial t = i[\cH,S(t)]$, to find
\begin{eqnarray}
\langle \dot{S}(\bk,t) \rangle =&&
-i\frac{\epsilon}{2}\tr\{e^{-i\cH t}[\cH,S(-\bk,0)]e^{i\cH
t}S(\bk,0)\} \nonumber \\
=&& -\frac{\epsilon}{2}\tr\{\dot{S}(-\bk,0)S(\bk,t)\} \nonumber \\
=&& -\frac{\epsilon}{2}\int_0^t
dt'\,\tr\{\dot{S}(-\bk,0)\dot{S}(\bk,t')\} \nonumber\\
&& - \frac{\epsilon}{2}\tr\{\dot{S}(-\bk,0)S(\bk,0)\}.
\end{eqnarray}
The second term on the right-hand side of the last equation
vanishes. Substituting the continuity equation, $\dot{S}(\bk,t) +
i\bk\cdot\bj(\bk,t) = 0$, and taking $\bk = k\hat{z}$, we obtain
\begin{equation}
\langle \dot{S}(\bk,t) \rangle = - \frac{\epsilon}{2}k^2\int_0^t
dt'\,\tr\{j_z(\bk,t')j_z(-\bk,0)\}. \label{numerator}
\end{equation}
According to Eq.\ (\ref{eq}), we further have
\begin{equation}
\langle S(\bk,0) \rangle =
\frac{\epsilon}{2}\tr\{S(\bk,0)S(-\bk,0)\}. \label{denominator}
\end{equation}
Substituting Eqs.\ (\ref{numerator}) and (\ref{denominator}) into
Eq.\ (\ref{LGK}) gives
\begin{equation}
D = \lim_{k\rightarrow 0}
\frac{\int_0^tdt'\,\tr\{j_z(\bk,t')j_z(-\bk,0)\}}{\tr\{S(\bk,0)S(-\bk,0)\}}.
\end{equation}
Taking the limit $t \rightarrow \infty$, we obtain the standard
form of the Kubo formula, Eq.\ (\ref{Kubo1}), which proves the
equivalence of the two approaches.

\end{document}